\documentclass[journal,article,submit,pdftex,moreauthors]{Definitions/mdpi} 

\firstpage{1} 
\pubvolume{1}
\issuenum{1}
\articlenumber{0}
\pubyear{2022}
\copyrightyear{2022}
\datereceived{} 
\dateaccepted{} 
\datepublished{} 
\hreflink{https://doi.org/} 
\preto{\abstractkeywords}{\nolinenumbers}
\Title{A smooth transition towards a Tracy-Widom distribution for the largest eigenvalue of interacting $k$-body fermionic Embedded Gaussian Ensembles}

\TitleCitation{A smooth transition towards a Tracy-Widom distribution for the largest eigenvalue of interacting $k$-body fermionic Embedded Gaussian Ensembles}

\Author{Ernesto Carro $^{1,*}$\orcidA{}, Luis Benet $^{1}$\orcidB{} and Isaac Pérez Castillo $^{2}$\orcidC}

\AuthorNames{Ernesto Carro, Luis Benet and Isaac Pérez Castillo}

\AuthorCitation{Carro E.; Benet L.; Pérez Castillo I.}

\address{$^{1}$ \quad Instituto de Ciencias Físicas, Universidad Nacional Autónoma de México, Av. Universidad s/n, Col. Chamilpa, CP 62210 Cuernavaca, Mor., Mexico; jecarro@icf.unam.mx; benet@icf.unam.mx\\
$^{2}$ \quad Departamento de Física, Universidad Autónoma Metropolitana-Iztapalapa, 
 San Rafael Atlixco 186, Ciudad de México 09340, Mexico; iperez@izt.uam.mx}

\corres{Correspondence: jecarro@icf.unam.mx}

\abstract{In spite of its simplicity, the central limit theorem captures one of the most outstanding phenomena in mathematical physics, that of universality. It states that, while navigating the set of all possible distributions by their convolution there exists a fixed point, the family of  Gaussian distributions, with a fairly large basin of attraction comprising a wide family of weakly correlated distributions.  More colloquially, this means  roughly speaking that the sum of independent and identically distributed random variables follows a Gaussian distribution irrespective of the details of the original distributions from which these random variables are drawn. 
While this classical result is well understood, and is the reason behind many physical phenomena, it is still not very clear what happens to this universal behaviour when the random variables become correlated. Not a general overarching theorem exists on a possible emergence of new universal behaviour and, so far as we are aware of, this area of research must be explored in a case-by-case basis. A fruitful mathematically laboratory to investigate the rising of new universal properties is offered by the set of eigenvalues of random matrices. In this regard a lot of work has been done using the standard random matrix ensembles and focusing on the distribution of extreme eigenvalues. In this case, the distribution of the largest ---or smallest--- eigenvalue departs from the Fisher-Tippett-Gnedenko theorem yielding the celebrated Tracy-Widom distribution. 
One may wonder, yet again, how robust is this new universal behaviour captured by the Tracy-Widom distribution when the correlation among eigenvalues changes. Few answers have been provided to this poignant question and our intention in the present work is to contribute to this interesting unexplored territory. Thus, we study numerically the probability distribution for the normalized largest eigenvalue of the interacting $k$-body fermionic orthogonal and unitary Embedded Gaussian Ensembles in the diluted limit. We find a smooth transition from a slightly asymmetric Gaussian-like distribution, for small $k/m$, to the Tracy-Widom distribution as $k/m\to 1$, where $k$ is the rank of the interaction and $m$ is the number of fermions. Correlations at the edge of the spectrum are stronger for small values of $k/m$, and are independent of the number of particles considered.  Our results indicate that subtle correlations towards the edge of the spectrum distinguish the statistical properties of the spectrum of interacting many-body systems in the dilute limit, from those expected for the standard random matrix ensembles.}


\begin{document}

\maketitle
\section{Introduction}
An important goal of basic sciences is to boldly search for universal behaviour, a common denominator that underlies the description of phenomena based on simple ingredients, either in the physical or mathematical reality. When this occurs they tend to be a paradigm-shifting moment:  Newton's, Maxwell's, Einstein's, Boltzmann's seminal works in physics come running to mind as well as those of Descartes, Fermat, Klein, Hamilton, Riemann, Langlands, and many other seminal works in the realm of mathematics. 
Focusing on the physical reality, whatever that means, we all would agree that many, if not all, physical systems can be modelled as a set of interacting random variables. This is apparent when describing the macroscopic behaviour of systems composed of a large number of constituents. Here the main goal of statistical mechanics is, by relating the macrostate of a system with all the microstates compatible to it, to explain all phases of matter and transitions between them. While descriptions of pure phases rely on the central limit theorem, close to a phase transition, when the microscopic constituents of the system become more and more correlated, the law of large numbers miserably fails and other techniques, such a renormalization group, must be used in its stead. 
Obviously, studying one physical system after another, identifying their basic constituents and their interactions, and seeking suitable mathematical and physical techniques to analyse them can be a tall order and, at times, quite frankly, tiring. In the last couple of decades, a more mundane and basic approach has been used: look for those mathematical models in which correlations of random variables can be easily modelled and manipulated, and study in these systems the emergence of new universal laws. Due to this, Random Matrix Theory (RMT) has been at the forefront in the study of emergent behaviour of correlated random variables.
Originally introduced to deal with the complexities of heavy nuclei Hamiltonian systems \cite{wign51} ---and historically also to deal with noisy linear systems of equations \cite{von1947numerical}--- RMT has grown to be an extremely successful theory with a surprisingly wide range of applications \cite{brody1981random,guhr98,RevModPhys69731,laloux2000random,luo2006application,tulino2004random,akemann2011oxford}. Its three main symmetry classes, the so-called canonical ensembles, were originally unveiled in \cite{dyso63, mehta} and several others \cite{forrester2010log} ---like the Wishart \cite{wishart1928generalised}, circular, or non-Hermitian ensembles--- took more relevance over the years.

It was first Tracy and Widom~\cite{trac93,trac94,trac96} who dared to look at how the probability distribution of the largest eigenvalue typically behaves, and whether its distribution departed from the one described by the extreme value theorem of Fisher-Tippett-Gnedenko for independent and identically distributed random variables \cite{fisher1928limiting,gnedenko1943distribution}. It was noticed that a new emergent distribution appears, the now celebrated Tracy-Widom distribution. After this, a flurry of research followed suit, originated by the seminal work of Dean and Majumdar~\cite{maj06}, focusing primarily in understanding the large deviation properties of extreme eigenvalues in standard ensembles of random matrices. By exploiting Dyson's log-gas analogy and, shrewdly using saddle-point techniques, they managed to obtain the left and right rate functions of extreme eigenvalues. Importantly, they noticed that the deviations to the left of, say, the largest eigenvalues are markedly different to the ones on its right, scaling differently with the system size. More research was done along these lines for other standard random matrix ensembles~\cite{PhysRevE.77.041108,PhysRevLett.102.060601,PhysRevLett.103.220603,PhysRevE.82.040104,Ramli_2012,PhysRevE.90.040102,PhysRevE.90.050103,castillo2016large,PhysRevE.100.012137, PhysRevLett.119.060601}, in diluted ensembles of random matrices \cite{PhysRevLett.117.104101,PhysRevB.96.064202,Lacroix_A_Chez_Toine_2018,PhysRevE.97.032124,PhysRevE.98.020102,castillo2021analytic}, and generalizations based on Dyson's log-gas analogy~\cite{PhysRevE.98.020104, Flack_2021, Flack_2022}, among many others, to ascertain the robustness of this new emergent law on the statistics of extreme values.

While RMT has been a fruitful mathematical laboratory in this particular endeavour, it was also recognized that its ensembles are somewhat unrealistic in the sense that they assume interactions to involve all Hilbert-space states, whereas typical forces in nature involve two-, three-, or a few-body interactions. This criticism led to the introduction of the two-body ensembles~\cite{fre70, boh71a, fre71, boh71b}, which eventually was generalized to the $k$-body embedded ensembles by French and Mon \cite{mon75}. Many results are known for the $k$-body embedded ensembles of Gaussian random matrices, which mostly focus on the properties of the mean-level density~\cite{BRW01, bene03, Kota2014}. Regarding the fluctuation properties of the spectrum, while it has not been proven that  they are of RMT type for the fermionic two-body embedded ensembles in the limit of large matrices,  there is vast numerical evidence that points on that direction, at least for the central part of the spectrum; see  e.g.~\cite{Papenbrock2006,Papenbrock2011,Kota2014}. As far as we are aware of, much less is known about the behavior at the edge of the spectrum for these ensembles. The main goal of the present paper is to  address the fluctuation properties of the largest eigenvalue of the fermionic $k$-body embedded ensembles of Gaussian random matrices in terms of its parameters. As, in contrast to the standard matrix ensembles, the number of mathematical techniques to tackle statistical problems in embedded ensembles is rather limited, our analysis is solely based in numerical methods.

\section{Matrix model and definitions}
Recall that the $k$-body fermionic Embedded Gaussian Ensemble (fEGE) is  the family of  matrix representations of  quantum Hamiltonian systems consisting of $m$ interacting spinless fermions, which can occupy any of $\ell$ possible degenerate single-particle states. The interaction among them is taken to be a $k$-body operator. More concretely, following \cite{bene03}, we introduce the operator $\Psi_{k; \rho}^\dagger \equiv \Psi_{j_1\dots j_k}^\dagger = \prod_{s=1}^k a_{j_s}^\dagger$ that creates $k\le m$ particles.  Here, $a_{j_s}^\dagger$ is the fermionic creation operator of the single-particle state $j_s$, while $\rho$ is the set of indices $(j_1, j_2, \dots, j_k)$, with the ordering $1\le j_1<\dots<j_k\le \ell$. In the number operator representation, the $k$-body interaction is then given  by
\begin{equation}
    \label{eq:Ham}
    V_{k}^{(\beta)} = \sum_{\rho,\sigma} v_{k; \rho,\sigma}^{(\beta)} \Psi_{k; \rho}^\dagger \Psi_{k; \sigma},
\end{equation}
where the coefficients $v_{k; \rho,\sigma}^{(\beta)}$, while obeying that $v_{k; \rho,\sigma}^{(\beta)}=[v_{k;\sigma,\rho}^{(\beta)}]^\star$, are independently distributed Gaussian random variables with zero mean and a constant variance, which henceforth is fixed to one. Finally, depending on Dyson's $\beta$ parameter, the set of coefficients $v_{k; \rho,\sigma}^{(\beta)}$ is either real (for $\beta=1$) or complex (for $\beta=2$). We refer to $k$ in Eq.~(\ref{eq:Ham}) as the rank of the interaction. The $m$-particle Hilbert space is spanned by a basis created by filling up $m$ states out of $\ell$, that is $|\mu\rangle = \Psi_{m;\mu}^\dagger|0\rangle$, which readily implies that this space has dimension  $N=\binom{\ell}{m}$. Using this basis, the matrix representation of the $k$-body interaction has entries given by $\langle\mu | V_k^{(\beta)} | \nu\rangle = \langle 0 | \Psi_{m;\mu} V_k^{(\beta)} \Psi_{m;\nu}^\dagger | 0 \rangle$. The particular case of $k=m$ coincides with the classical Gaussian ensembles of RMT, while for $m>k$ the matrix elements $\langle\mu | V_k^{(\beta)} | \nu\rangle$ display correlations or may even be  identically zero \cite{bene03}.

The limit $N\to\infty$ for the fEGEs can be attained either as $\ell\to\infty$ for fixed $m$, or by fixing the filling factor $m/\ell$ and taking the limits $m,\ell\to\infty$. These limits have different properties in terms of the moments of the density of states~\cite{BRW01,Kota2014}. In the present study, we shall focus on the limit where $m$ remains constant, which is more advantageous from the numerical point of view. This is the so-called dilute limit.

To characterize the probability distribution of the largest eigenvalue for the fEGEs, that we denote as $\lambda_\textrm{N}$, we assume that its mean value can be written as $\langle\lambda_\textrm{N}\rangle = \sqrt{2\beta}N^\alpha$, while its standard deviation behaves as $\sigma_{\lambda_{\textrm{N}}}\sim N^\gamma$. Here $\langle(\cdots)\rangle$ corresponds the the ensemble average in the embedded ensemble. The normalized largest eigenvalue, denoted as $\tilde{\lambda}_\textrm{N}$, is then given by
\begin{equation}
    \label{eq:lambda}
    \tilde{\lambda}_\textrm{N} = \frac{\lambda_\textrm{N} - \sqrt{2\beta} N^{\alpha}}{N^\gamma}.
\end{equation}
As it is well known, for the standard Gaussian ensembles, which correspond in our case $k=m$, the celebrated Tracy-Widom distribution corresponds to having scaling exponents  $\alpha=1/2$ and  $\gamma=-1/6$ \cite{trac93,trac94,trac96}. 

\section{Results}
For the fEGEs the scaling exponents remain unknown though. As an exact mathematical analysis is rather difficult, we henceforth rely on numerical methods in order to infer $\alpha$ and $\gamma$ in terms of the parameters defining the EGEs. We use a simple random sampling method to generate $10^4$ matrices belonging to the fEGEs for both $\beta=1$ and $\beta=2$ and for different number of fermions (parameter $m$) and single-particle states (controlled by the parameter $\ell$). More concretely, we take $m=4$ with values of $\ell\in\{8,9,\ldots, 28\}$, $m=5$ with $\ell\in\{10,11,\ldots, 20\}$, $m=6$ and $\ell\in\{12,13,\ldots, 18\}$, $m=7$ with $\ell\in\{14,15,\ldots, 18\}$, and, finally, $m=8$ with values $\ell\in\{16,\ldots, 18\}$. Besides, in all cases the rank takes values $k=1,\ldots,m$.  Each choice of $m$ and $\ell$ fixes the linear dimension $N$ of the matrix, and by varying $\ell$ we can see how the scaling exponents behave with $N$. This is then repeated for different ranks $k$ of the interaction. 

\begin{figure}
  \includegraphics[width=0.6\textwidth]{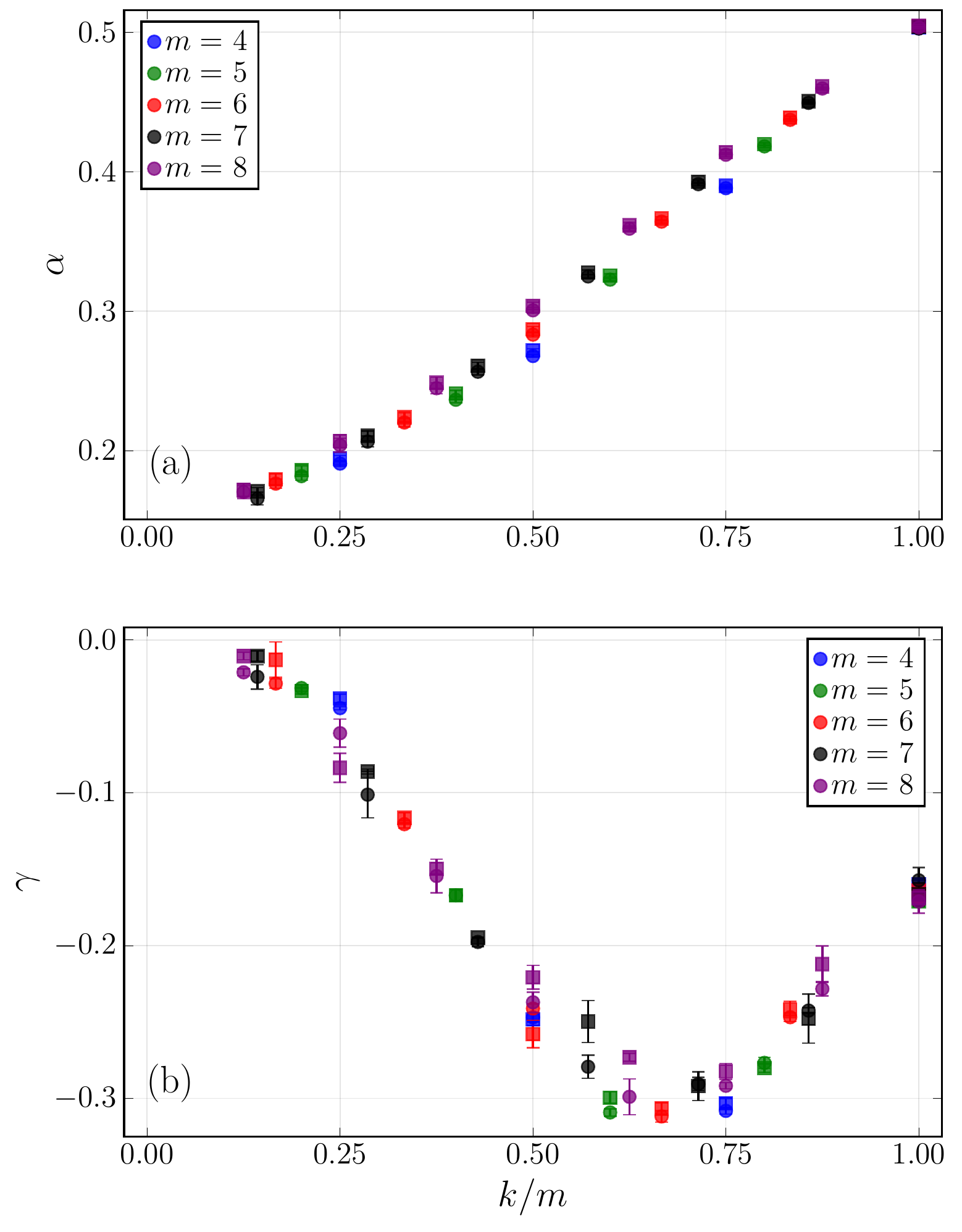}
  \caption{Scaling exponents $\alpha$ and $\gamma$ for the  mean and standard deviation of $\lambda_\textrm{N}$ for the  fEGEs, in terms of $k/m$. Data points represented by circles correspond  to the orthogonal ensembles ($\beta=1$) while data for the unitary  case ($\beta=2$) is represented by squares.}
  \label{fig1}
\end{figure}

Figure \ref{fig1} summarises the results for $\alpha$ (top figure) and $\gamma$ (bottom figure) as a function of the ratio  $k/m$ for $\beta=1$ (represented by circles) and  $\beta=2$  (represented by squares in the same figure). The error bars are the result of estimating the exponents by linear regression. As it can be appreciated, both exponents, when plotted as a function of $k/m$, seem to coalesce into a curve, independent on the value of $\beta$. While the exponent $\alpha$ is monotonously increasing, the exponent $\gamma$ that controls the amplitude of the fluctuations has, rather surprisingly, a minimum at around $k/m\sim 0.7$. Moreover, when $k/m=1$ we recover the expected values of both exponents according to the Tracy-Widom distribution.

\begin{figure}
    \includegraphics[width=0.6\textwidth]{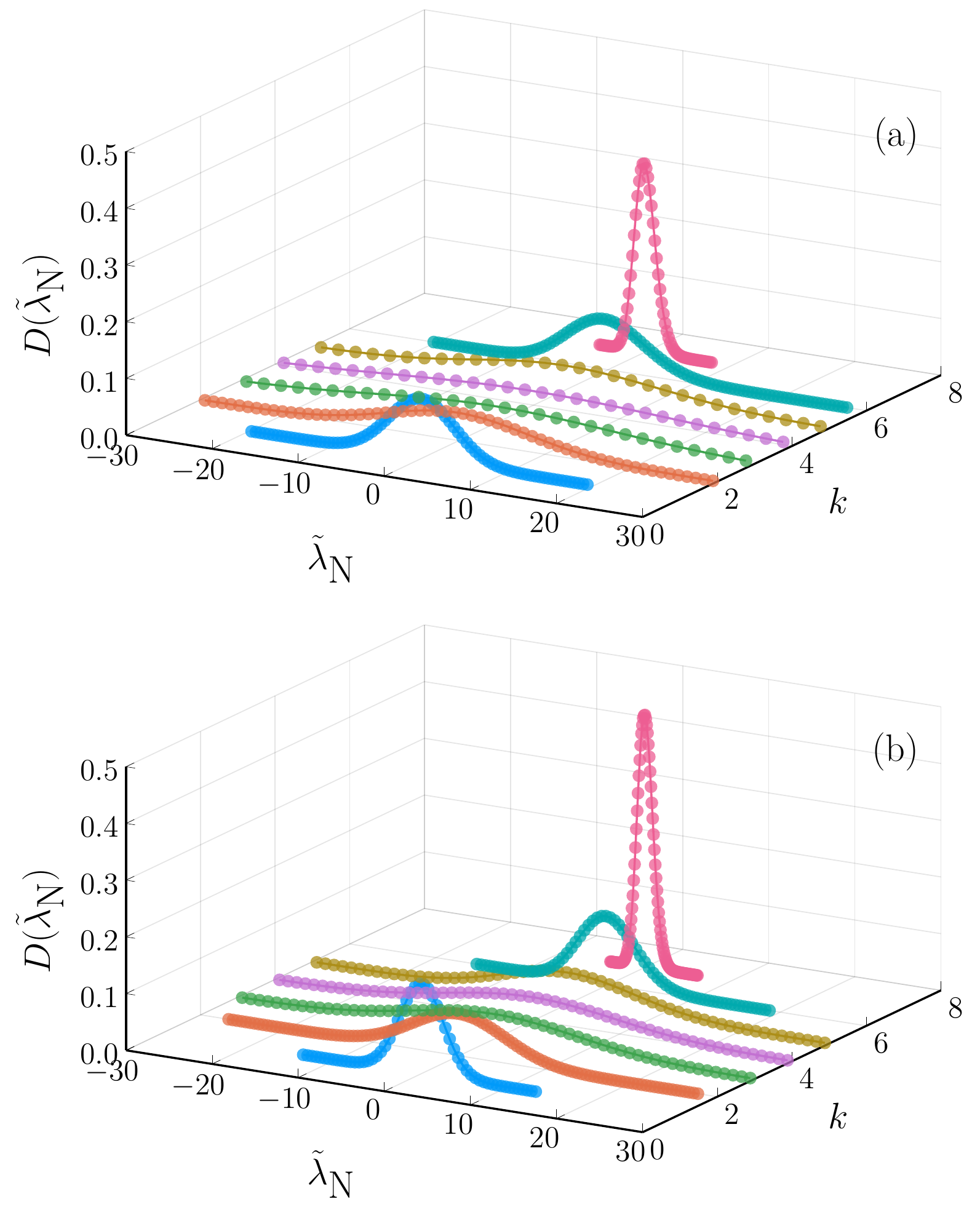}
    \caption[]{Probability distribution $D(\tilde{\lambda}_\textrm{N})$ for the normalized largest eigenvalue in terms of $k$, for $m=7$ and $\ell = 14$ ($f = 1/2$) for (a)$\beta=1$ and (b)~$\beta=2$. The resulting distributions display a transition 
    from an almost-symmetric Gaussian distribution for $k=1$, to the Tracy-Widom distribution for $k=m$.}
    \label{fig2}
\end{figure}

\begin{figure}
  \includegraphics[width=0.6\textwidth]{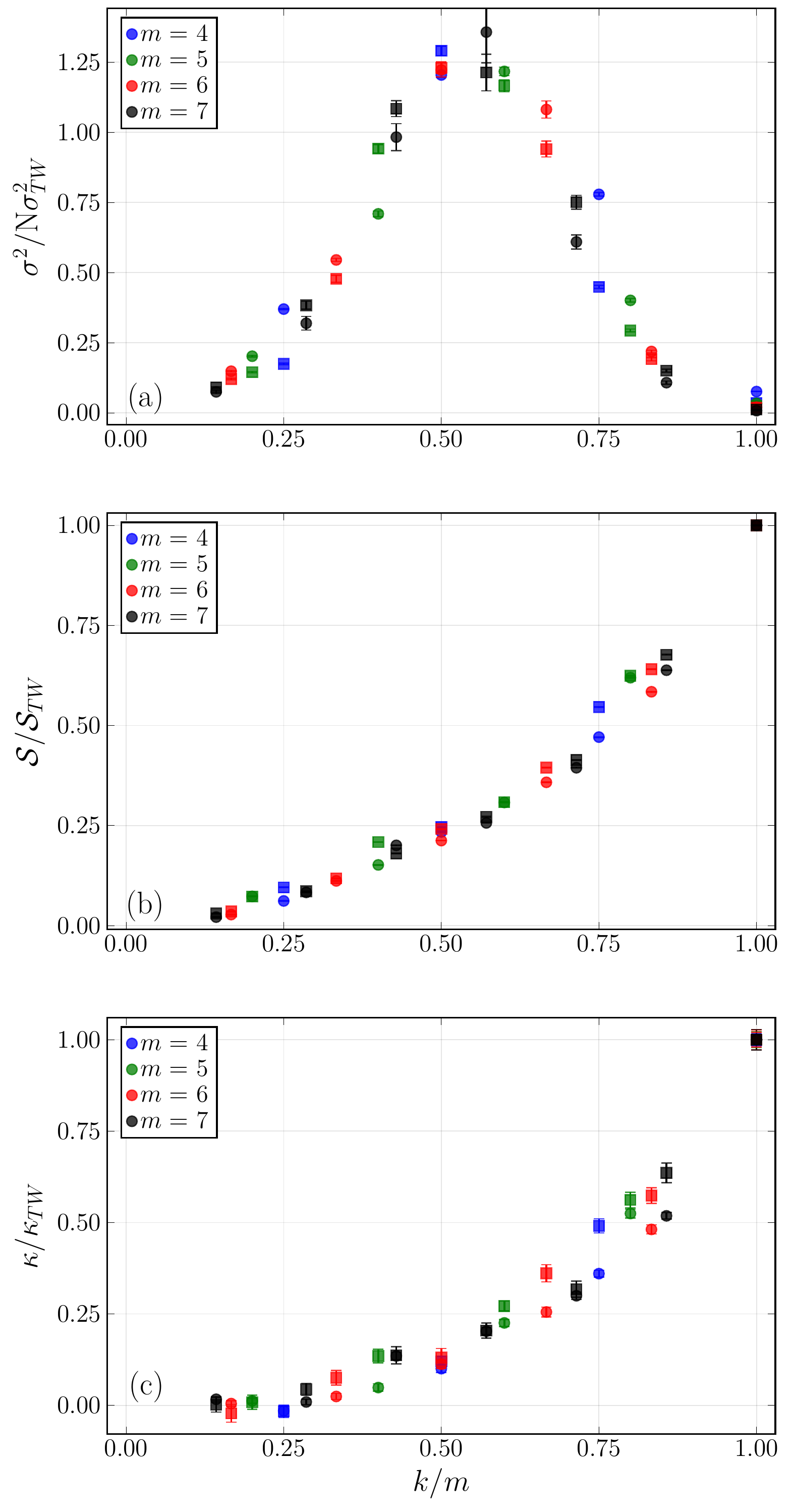}
  \caption[]{(a) Variance divided by $N$, (b)~skewness and (c)~excess kurtosis, normalized to the corresponding values of the Tracy-Widom distribution, for $D(\tilde{\lambda}_\textrm{N})$ in terms of $k/m$, for $\ell=14$. Circles correspond to the results obtained for $\beta=1$
  and squares to the case $\beta=2$.}
\label{fig3}
\end{figure}

To dive deeper into the probability density function of the largest eigenvalue, denoted as $D(\tilde{\lambda}_\textrm{N})$, let us take, for sake of discussion, the values of $m=7$ and $\ell=14$, and investigate how the shape of the distribution changes as we vary the rank of the interaction. This is precisely shown in Figure~\ref{fig2}; the top figure corresponds to $\beta=1$ while the bottom one is for the unitary case $\beta=2$. To quantify better the profile of $D(\tilde{\lambda}_\textrm{N})$, in Figure~\ref{fig3} we also show the behaviour of its variance, skewness and excess kurtosis, normalised with respect to the ones of the Tracy-Widom, as a function of $k/m$. In the diluted limit we are working on, we observe a smooth transition from an almost symmetric Gaussian distribution for small values of $k/m$, to a more spread and asymmetric distribution as $k/m$ is increased, to finally arriving to the Tracy-Widom distribution for $k/m=1$. Moreover, at around $k\simeq m/2$ the width of the distribution ceases to grow. These results indicate that the correlations at the edge of the spectrum are different in terms of $k/m$. While this statement is obviously for $k/m=1$, our findings for the one-body interaction $k=1$ are actually rather surprising. Here, according to \cite{guhr98} or \cite{BRW01}, we would expect the spectral statistics in the bulk to correspond to an uncorrelated Poisson distribution which should na\"ively yield, in turn, to an extreme value distribution according to the Fisher-Tippett-Gnedenko theorem \cite{Fortin2015,Vivo2015,Hansen2020}. More precisely, since for $k=1$ the mean-level density is Gaussian \cite{fre70,BRW01,bene03,Kota2014}, an uncorrelated spectrum should yield a Gumbel distribution \cite{Bouchaud1997, Fortin2015,Vivo2015,Hansen2020}, but this is not the result obtained for $k=1$. Instead, our distribution is closed to a slightly asymmetric Gaussian distribution. Thus, there sill must be important correlations occurring towards the edge of the spectrum also in this case.

To unveil this oddity, we proceed to estimate the correlation coefficient between the largest and second largest eigenvalues considering the formula:
\begin{eqnarray}
    \label{eq-corr}
    c_{\textrm{N}, \textrm{N-1}} & = & \langle\tilde{\lambda}_\textrm{N} \tilde{\lambda}_\textrm{N-1}\rangle - \langle\tilde{\lambda}_\textrm{N}\rangle \langle\tilde{\lambda}_\textrm{N-1}\rangle\,.
\end{eqnarray}
In addition, we shall also estimate the distribution of the normalized distance $s = (\lambda_\textrm{N}-\lambda_\textrm{N-1})/\langle\lambda_\textrm{N}-\lambda_\textrm{N-1}\rangle$ between these two  eigenvalues.

\begin{figure}
  \includegraphics[width=0.6\textwidth]{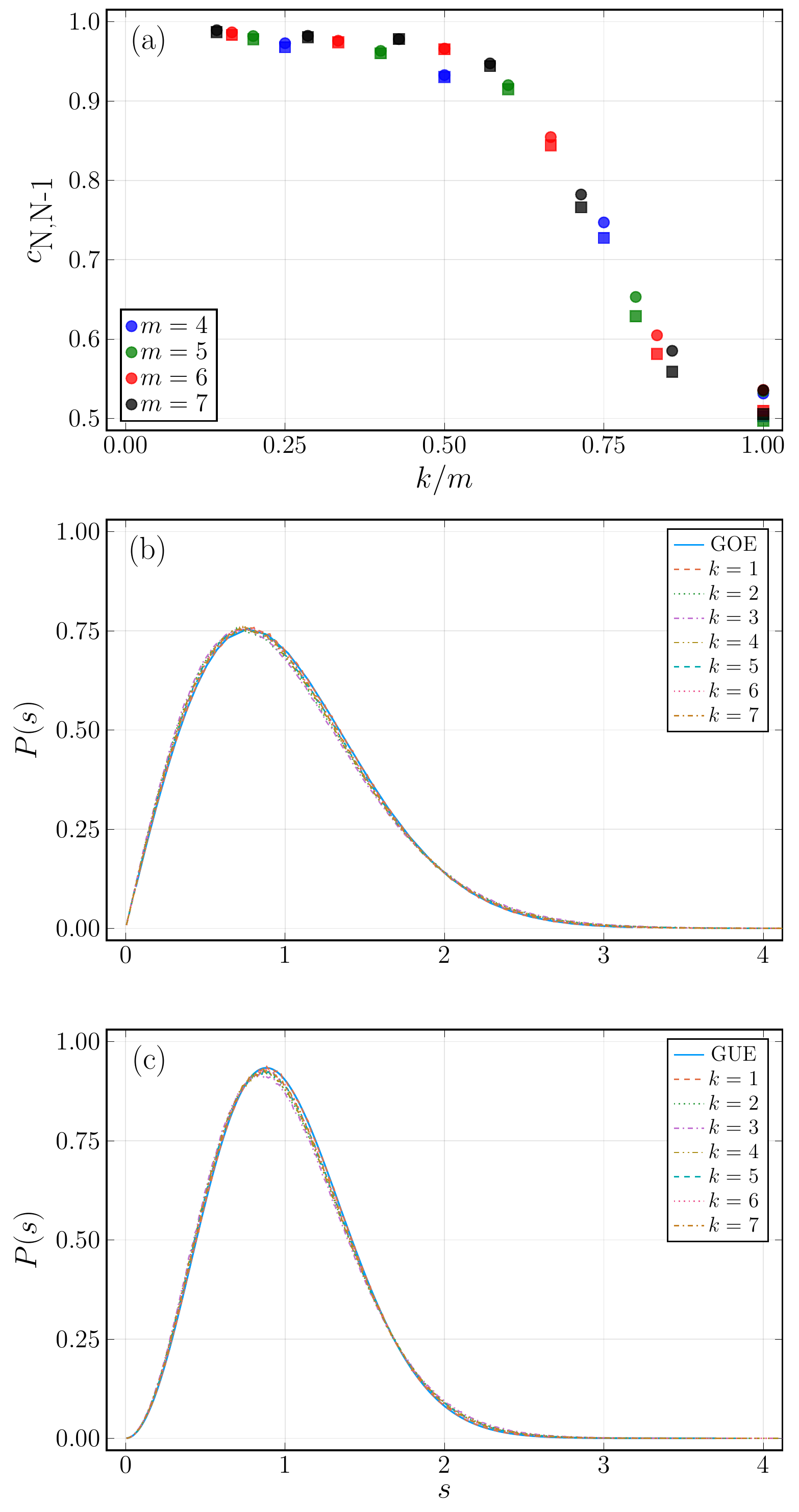}
  \caption[]{(a) Normalized correlation coefficient, Eq.~(\ref{eq-corr}), for the two largest eigenvalues of the 
  fEGEs in terms of $k/m$ with $\ell=14$, for $\beta=1$ (circles) and $\beta=2$. (b) Distribution $P(s)$ of the (normalized) distance among the  largest eigenvalues for $m=7$, $\ell=14$, and $\beta=1$, and  (c)~$\beta=2$. The continuous curves correspond to the nearest-neighbor distribution obtained by the Wigner surmise 
  for the RMT ensembles.} 
  \label{fig4}
\end{figure}

In Figure~\ref{fig4}(a), that shows  the correlation coefficient $c_{\textrm{N}, \textrm{N-1}}$, we notice it is a positive and monotonically decreasing function of $k/m$. Besides, its behavior seems to be independent of Dyson's index. Fairly interestingly, for small values of $k/m$, and in particular for $k=1$, the correlations are essentially twice stronger compare to the Tracy-Widom case of $k=m$. Moreover, the correlation coefficient seems to have a plateau ---or to decay rather slowly--- for values of $k/m\leq 1/2$, above which it starts to decay smoothly but rapidly towards the value predicted by the standard RMT ensembles.

In Figure~\ref{fig4}(b) and~(c) we present the results for the distribution of the normalized distance $s$ among the largest eigenvalues, for the orthogonal and unitary fEGEs, respectively, and various values of $k$, for $m=7$ and $\ell=14$. This quantity is analogous to the nearest-neighbour distribution used widely to characterize the spectral fluctuations at the bulk of the spectrum~\cite{guhr98}. We observe that the distributions obtained are weakly dependent on the rank of the interaction $k$, and depend on $\beta$ exhibiting the usual level repulsion $s^\beta$. It is tempting to compare these distributions with the $P(s)$ obtained for the canonical ensembles of RMT~\cite{DietzHaake1990}, as it has been done in the past for the two-body random ensembles ($\beta=1$)~\cite{boh71b}; we note that the results for the two-body random ensembles were compared with the Wigner surmise~\cite{boh71b}. The results indicate that these distributions are quite close to the nearest-neighbour distribution obtained for the corresponding RMT ensembles; we attribute the weak differences observed in terms of $k$ as finite size effects. To be more quantitative, we computed the mean square-differences of the numerically obtained distributions for the fEGEs and the corresponding $P(s)$ for the RMT ensembles. The results for $m=7$ and $\ell=14$ show that mean square-differences from $k=1$ grow, reaching a maximum value at $k=3$, and then decrease; small quantitative differences appear in terms of $\beta$. Interestingly again, the minimum mean square-difference is attained at $k=1$ for both values of $\beta$.

\section{Summary and conclusions}
In this paper we have studied the statistical  properties of the largest eigenvalue for the orthogonal and unitary $k$-body interacting fEGEs. We have presented numerical evidence showing that, in the dilute limit, there is a smooth transition in the distribution of the largest eigenvalue from a slightly asymmetric Gaussian-like distribution for small values of $k/m$, to the Tracy-Widom distribution as $k/m$ approaches one. This transition is such that both the normalized (with respect to the Tracy-Widom corresponding quantities) skewness and the excess kurtosis grow smoothly with respect to $k/m$,  while the normalized width of the distribution displays a maximum and is somewhat asymmetric with respect to $k/m$. Based on these findings, it is clear that for $k\ll m$ the the distribution of the largest normalized eigenvalue are different from the Tracy-Widom distribution~\cite{DeanEtal2016}: while it is difficult to attain numerically the case $k/m \to 0$ in the dilute limit, our results do indicate that the distribution of the largest normalized eigenvalue is close to a Gaussian, perhaps reaching the Gaussian limit in the large $N$ limit. We also studied the correlation coefficient between the largest and second largest eigenvalues finding it to be a strictly-positive monotonically decreasing function of $k/m$, with different decay rates depending on the value $k/m$. Yet, the distribution of the distance among the two largest eigenvalues of the fEGEs seems to be quite close to the nearest-neighbor distribution of the canonical RMT ensembles in the bulk, with a weak dependence on $k$. These results show that, while the statistics at the bulk of the spectrum for the fEGEs may coincide with those of the canonical ensembles, say for $k=2$, correlations that depend on $k$ arise towards the edge of the spectrum. The properties of the distributions of the largest eigenvalue for the fEGEs in the limit $N\to\infty$ with constant filling factor remain an open problem as well as an analytical confirmation of our findings.

\vspace{6pt} 
\authorcontributions{Conceptualization and methodology, L.B. and I.P.C.;  software, L.B. and E.C.; validation, E.C; all authors have contributed to writing and editing. All authors have read and agreed to the published version of the manuscript.}

\funding{This research was funded by UNAM--PAPIIT grant number IG-101122.}


\acknowledgments{L.B. is thankful to François Leyvraz and Hernán Larralde for illuminating discussions. E.C. acknowledges a doctoral fellowship provided by CONACyT.}

\conflictsofinterest{The authors declare no conflict of interest. }

\begin{adjustwidth}{-\extralength}{0cm}
\reftitle{References}
\bibliography{my_bib.bib}

\end{adjustwidth}
\end{document}